\documentclass{article}
\usepackage{acra}
\usepackage{amsmath,amssymb,amsfonts}
\usepackage{algorithmic}
\usepackage{graphicx,url}
\usepackage{textcomp}
\usepackage{xcolor,bm}
\def\BibTeX{{\rm B\kern-.05em{\sc i\kern-.025em b}\kern-.08em
    T\kern-.1667em\lower.7ex\hbox{E}\kern-.125emX}}
\usepackage{subfiles} 
\graphicspath{ {./Figures/} }
\newcommand\numberthis{\addtocounter{equation}{1}\tag{\theequation}}
\usepackage{comment}
\usepackage{subcaption}
\usepackage{float}
\title{FTISS Adaptive Bearing-Only Formation Tracking Control with Unknown Disturbance Rejection}
\author{Hong Liang Cheah and Mohammad Deghat\\School of Mechanical and Manufacturing Engineering\\ University of New South Wales, Australia \\ 
\{h.cheah, m.deghat\}@unsw.edu.au}

\begin{document}

\maketitle

\begin{abstract}
This paper proposes a finite-time input-to-state stable (FTISS) bearing-only formation control law that rejects unknown constant disturbances. Unlike existing finite-time bearing-based formation control laws, which typically rely on the availability of a global coordinate frame and some information about the disturbances, our approach requires only local bearing vector measurements and does not necessitate the alignment of agent coordinate frames. The proposed control law guarantees that formation control errors converge to a neighborhood of zero in finite time, and subsequently converge to zero asymptotically. We first address the scenario where leaders are stationary and then extend the results to leaders moving with a constant velocity. Simulation and experimental results are presented to validate the effectiveness of the proposed control law.
\end{abstract}

\section{Introduction}

The existing multi-agent formation control techniques can be categorized into three main methods: position-based, distance-based, and bearing-based. Among these techniques, bearing-based formation control has recently attracted significant attention from the control community. This is due to the easily obtainable relative bearing measurements using sensor arrays \cite{mao_2007_wireless} or lightweight cameras \cite{tron_2016_a}, compared to distance-based formation control, where relative position measurements are usually required. The objective of bearing-based formation control is to control a group of agents to achieve a desired formation configuration, defined by relative bearing vectors or bearing angles.


Bearing rigidity theory, proposed in \cite{zhao_2016_bearing}, characterizes unique formation shapes from bearing vector constraints. Building upon this, bearing-based leader-follower formation control was studied in \cite{zhao_2016_localizability,zhao_2015_maneuver,VanVu_2021_decentralized}. However, these control laws require agents to share a common coordinate frame as the desired formation is defined by coordinate-dependent bearing vectors. In the absence of a common coordinate frame among agents, orientation synchronization or estimation techniques are often employed to achieve consensus on the coordinate frame \cite{zhao_2016_bearing,vantran_2019_finitetime,xu_2023_bearingbased}. However, these methods require agents to communicate with their neighbors to exchange bearing angle measurements \cite{Ahn_2020_formation}. 


\begin{figure}[t] 
    \centering
    \includegraphics[trim={0 25mm 0 10mm}, clip,width=0.9\linewidth]{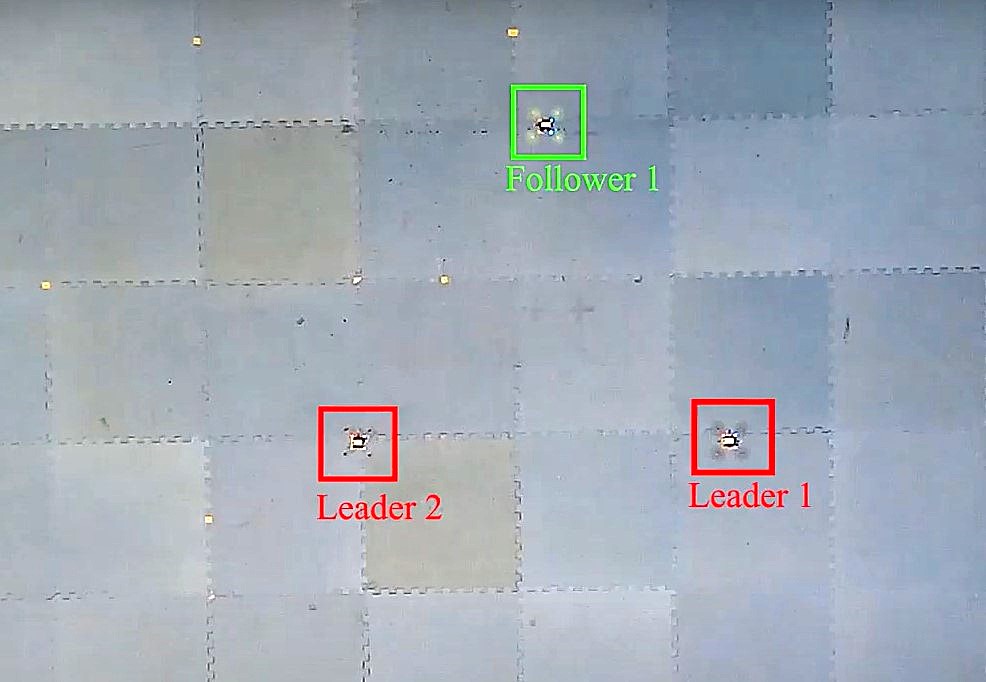}
    \caption{Formation control of three drones arranged in a triangular configuration.}
    \label{exss}
\end{figure}

Communication in formation control is often undesirable because of the potential for delays and packet losses. As a result, formation control using bearing angles has been explored in, for example, \cite{chen_2021_angle,Jing_2019_angle,Bishop_2015_distributed}. Since the desired formation is defined by scalar bearing angles, a global coordinate frame is not required. However, such formation control laws are only applicable to planar formations. Although 3D angle-based formation control was studied in
\cite{chen_2023_angle,chen_2024_air}, it requires communication between agents to exchange angle measurements. An alternative approach, elevation angle-based formation control, was explored in \cite{chen_2022_gradientbased}, demonstrating its applicability to 2D and 3D formations while eliminating the need for communication between neighboring agents. In our previous work \cite{cheah_2024_bearingbased}, we studied leader-follower elevation angle-based formation tracking control for single-integrator, double-integrator, and non-holonomic robots. However, these elevation angle-based methods do not account for unknown disturbances.

Disturbances, such as wind, can affect the performance of formation control laws. Several studies have been dedicated to address disturbances in bearing-based formation control, see e.g., \cite{bae_2022_distributed,wu_2023_finitetime,wang_2022_adaptive,ding_2023_event}. However, these results only establish input-to-state stability, indicating that while bearing errors remain bounded, they do not necessarily converge to zero. Thus, to eliminate the effect of disturbances, several control laws have been proposed, and bearing errors successfully converged to zero. Specifically, to reject constant disturbances, a PI control law and an adaptive control law that guarantee asymptotic convergence were introduced in \cite{zhao_2017_translational} and \cite{chen_2022_fuzzy}. Robust bearing-based formation control laws have also been proposed in \cite{trinh_2021_robust,Wu_2024_distributed,vantran_2023_robust,song_2024_bearing} to reject time-varying disturbances, assuming that agents know an upper bound of disturbances. Furthermore, finite-time robust formation control with disturbance rejection has been studied for bearing-based formations in \cite{xu_2023_bearingbased}. Compared to controllers that only ensure asymptotic convergence, finite-time controllers deliver better transient responses \cite{bhat_2000_finitetime}. However, this control law works under the assumption that an upper bound on the disturbance is known and velocity measurements are required. Moreover, to address time-varying disturbances with an unknown upper bound, bearing-based robust-adaptive controllers were proposed in \cite{wang_2023_bearing,CHENG_2024_adaptive,nguyen_2023_bearingconstrained}, although these control laws only ensure asymptotic stability. An alternative robust-adaptive formation control law was proposed by \cite{Su_2024_bearing}, but it requires communication between neighboring agents. Nevertheless, the use of the discontinuous signum function in both robust or robust-adaptive formation control laws may induce chattering in control systems. Additionally, the aforementioned strategies require a global coordinate frame, since the desired formation is defined by coordinate-dependent bearing vectors. As a result, we aim to develop a novel finite-time control law that improves on the existing literature by rejecting unknown constant disturbances in agents' local coordinate frames.

When a global coordinate frame is not available, a leaderless elevation angle-based robust adaptive control law was adopted in \cite{Garanayak_2023_Bearingdisturbance}. However, the desired formation achieved will shift due to the presence of disturbances. This paper extends beyond the results presented in \cite{Garanayak_2023_Bearingdisturbance} and includes agents with a leader-follower structure. By incorporating leaders into the multi-agent system, the agents maintain the desired formation shape and do not move in the direction of the disturbance. 

This paper develops a control law with finite-time convergence properties and unknown constant disturbance rejection. This ensures that the formation error converges to a neighborhood of the equilibrium in finite time and then asymptotically converges to the equilibrium. The contributions of this work are summarized as follows:\\
($i$) We propose a novel FTISS bearing-only formation tracking control law that effectively rejects unknown constant disturbances without necessitating a global coordinate frame. Compared to the existing finite-time robust disturbance rejection formation control laws \cite{xu_2023_bearingbased,VanVu_2021_distance,cheah_2024_bearingbased}, the proposed method ensures that formation errors converge to a neighborhood of zero within a finite time, while handling unknown constant disturbances without requiring prior knowledge of their upper bound. In contrast to the adaptive distance-based formation control law in \cite{trinh_2022_adaptive}, the proposed control law ensures finite-time convergence to a neighborhood of zero and requires only local bearing measurements.  \\ 
($ii$) Unlike \cite{Garanayak_2023_Bearingdisturbance,bae_2022_distributed,wu_2023_finitetime,Wu_2024_distributed}, where only static formation is studied, we demonstrate that the proposed FTISS formation control law is also applicable to moving leaders with an unknown constant velocity. Furthermore, the proposed control law does not require communication between neighboring agents and relies exclusively on local bearing measurements.\\
($iii$) In contrast to many existing studies that solely present simulation results, we also provide experimental tests to demonstrate the feasibility of the proposed control law in real-world environments as shown in Figure \ref{exss}.  \vspace{0.2cm}\\
\textit{Notations:}\\
Denote $\mathbb{R}^+$ as the set of non-negative real numbers. Let $\bm{0}_d = [0,...,0]^\top\in\mathbb{R}^d$ and $\bm{1}_d = [1,...,1]^\top\in\mathbb{R}^d$. Denote the $d\times d$ zero matrix as $0_{d\times d}$. Let $\|\cdot\|$ represent the Euclidean norm of a vector or the spectral norm of a matrix, and let $\|\cdot\|_\infty$ be the infinity norm. The $d$-dimensional identity matrix is denoted by $I_d$, and $\otimes$ represents the Kronecker product. 
Let $\bm{x}=[x_1,...,x_d]^\top$ denote a column vector in $\mathbb{R}^d$. Consider two functions $\psi(\cdot)$ and $\bar{\psi}(\cdot)$: $\bar{\psi}(\bm{x})\sim\psi(\bm{x})$ means $c_1\psi(\bm{x})\leq\bar{\psi}(\bm{x})\leq c_2\psi(\bm{x})$ for some positive constants $c_1$ and $c_2$. For $\beta>0$, $|\bm{x}|^\beta$ represents $\sum_{i=1}^d |x_i|^\beta$, where $x_i$ is the $i$th element of $\bm{x}$. The function $\text{sig}(\cdot)^\beta:\mathbb{R}^d \rightarrow\mathbb{R}^d$ is defined as 
\begin{equation}\label{e1}
   \text{sig}(\bm{x})^\beta =[\text{sign}(x_1)|x_1|^\beta, ..., \text{sign}(x_d)|x_d|^\beta]^\top.    
\end{equation}
Note that the function $\text{sig}(\cdot)^\beta$ is continuous, odd, and 
\begin{equation} \label{signorm}
    \bm{x}^\top\text{sig}(\bm{x})^\beta = |\bm{x}|^{1+\beta}.
\end{equation}
Moreover, the following inequality holds
\begin{equation}\label{e2}
    |\bm{x}|^{1+\ell}= \sum_{i=1}^d |x_i|^{1+\ell} \geq \left(\sum_{i=1}^d |x_i|^2\right)^\frac{1+\ell}{2}=\|\bm{x}\|^{1+\ell}
\end{equation}
for $\ell\in(0,1)$, (see \cite[Lemma 3.3]{zuo_2014_anew}).

\section{Preliminaries}
\subsection{Graph Theory and Elevation Angle Rigidity}
Consider a group of $n$ agents in $\mathbb{R}^d$ ($n \geq 2$ and $d = 2$ or $3$). Let $\bm{p}_i(t) \in \mathbb{R}^d$ represent the position of agent $i \in {1, \ldots, n}$, and let $\bm{p}(t) = [\bm{p}_1(t)^\top, \ldots, \bm{p}_n(t)^\top]^\top \in \mathbb{R}^{dn}$ denote the configuration of the agents. The interaction among the agents is modeled by a fixed undirected graph $\mathcal{G} = (\mathcal{V}, \mathcal{E})$, where $\mathcal{V} = \{1, \ldots, n\}$ is the vertex set and $\mathcal{E} \subseteq \mathcal{V} \times \mathcal{V}$ is the edge set. The edge set $\mathcal{E}$ contains $m$ edges. An edge $(i,j) \in \mathcal{E}$ implies that agents $i$ and $j$ are neighbors, which means that they can measure the relative bearing of each other. The set of neighbors of agent $i$ is denoted by $\mathcal{N}_i = \{j \in \mathcal{V} : (i,j) \in \mathcal{E}\}$. A \textit{framework} is defined as a pair $(\mathcal{G}, \bm{p})$, where each vertex $i$ in $\mathcal{G}$ is mapped to $\bm{p}_i(t)$ for all $i \in \mathcal{V}$.

Let $Q_i \in \text{SO}(d)$ be the rotation matrix that rotates the local coordinate frame of agent $i$, which is a fixed frame known only to agent $i$, to the global coordinate frame, and let $\bm{T}_i \in \mathbb{R}^d$ represent the translation from the local frame of agent $i$ to the global coordinate frame. Note that both $Q_i$ and $\bm{T}_i$ are time-invariant, and that $Q_i^{-1} = Q_i^\top$ and $Q_i Q_i^\top = I_d$. Let the position of agent $i$ measured in its local coordinate frame be
\begin{equation}\label{e3}
    ^i\bm{p}_i(t)= Q_i^\top(\bm{p}_i(t)+\bm{T}_i).
\end{equation}
The edge vector $^i\bm{e}_{ij}$ and bearing vector $^i\bm{g}_{ij}$ measured in the local frame of agent $i$ are denoted as
\begin{equation}\label{e4}
    ^i\bm{e}_{ij}(t):={^i\bm{p}}_j(t)-{^i\bm{p}}_i(t),\ ^i\bm{g}_{ij}(t):=\frac{^i\bm{e}_{ij}(t)}{\| ^i\bm{e}_{ij}(t) \|},
\end{equation}
where ${^i\bm{p}}_j(t)$ is the position of neighbor $j$ measured in the coordinate frame of agent $i$. 
Note that $^i\bm{e}_{ij}(t) = -{^i\bm{e}}_{ji}(t) $ and $^i\bm{g}_{ij}(t) = -{^i\bm{g}}_{ji}(t)$. In the global coordinate frame, we have the edge vector $\bm{e}_{ij}(t)=Q_i {^i\bm{e}}_{ij}(t)$, and the bearing vector ${\bm{g}}_{ij}(t)=Q_i {^i\bm{g}}_{ij}(t)$. From now on, when no confusion arises, time dependencies in variables are omitted for ease of notation.

Consider an arbitrary orientation of graph $\mathcal{G}$. The incidence matrix $H \in \mathbb{R}^{m\times n}$ is a $\{0,\pm1\}$-matrix with rows indexed by edges and columns by vertices. The entries are defined as follows: $[H]_{ki} = 1$ if vertex $i$ is the head of edge $k$, $[H]_{ki} = -1$ if vertex $i$ is the tail of edge $k$, and $[H]_{ki} = 0$ otherwise. For a connected graph, it holds that $H\bm{1}_n = \bm{0}_m$ and $\text{rank}(H) = n-1$ \cite{Mesbahi_2010_graph}. 

\begin{figure}[thb!] 
    \centering
    \includegraphics[scale=1]{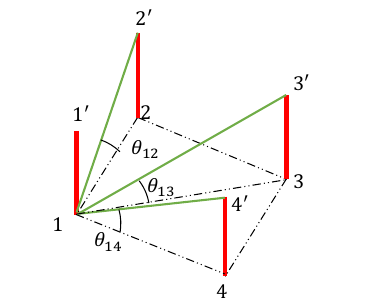}
    \caption{Elevation angles in 2D.}
    \label{f1}
\end{figure}

The idea of elevation angle rigidity was introduced in \cite{chen_2022_gradientbased}. In the 2D scenario, each agent is attached to a vertical rod of height $h_i > 0$, as illustrated in Figure \ref{f1}. This makes the position of the rod's endpoint $\bm{p}_{i'}=\bm{p}_i + [0, 0, h_i]^\top $, thus extending the 2D framework to 3D. The elevation angle $ \theta_{ij} \in (0, \frac{\pi}{2}) $ from agent $ i $ to agent $ j $ is defined as follows
\begin{equation} \label{e5}
    \theta_{ij} :=\angle jij'=\arccos(\bm{g}_{ij}^\top \bm{g}_{ij'})=\arctan(h_j/l_{ij}), 
\end{equation}
where $l_{ij}=\|^i\bm{e}_{ij}\|=\|\bm{e}_{ij}\|$ is the distance between agent $i$ and agent $j$. In this paper, it is assumed that the agent rods have the same height $h_i=h_c>0,\ \forall i \in\mathcal{V}$. Consequently, $\theta_{ij}=\theta_{ji}=\arctan(h_c/l_{ij})$. Figure \ref{f1} shows three elevation angles $\theta_{12},\ \theta_{13},\ \theta_{14}$ measured from agent 1. We define the elevation angle function $\bm{f_E}(\bm{p}):\mathbb{R}^{3n}\rightarrow \mathbb{R}^m$ as $\bm{f_E}(\bm{p}):=[f_1,...,f_k,...,f_m]^\top$, where 
\begin{equation} \label{e6}  
f_k:=f_{ij}=\cot(\theta_k)=\frac{\cos(\theta_{ij})}{\sin(\theta_{ij})}=\frac{l_{ij}}{h_{c}},\ k = 1,...,m.
\end{equation}
Note that $\theta_k = \theta_{ij}=\theta_{ji}$ is a scalar representing the elevation angle associated with the $k$th edge. To compute $f_{ij}$ using local bearing vectors, we have
\begin{equation}\label{e7}
    f_{ij}{=}\frac{{^i\bm{g}}_{ij}^\top {^i\bm{g}}_{ij'}}{\sqrt{1-({^i\bm{g}}_{ij}^\top {^i\bm{g}}_{ij'})^2}}\\
    = \frac{\bm{g}_{ij}^\top \bm{g}_{ij'}}{\sqrt{1-(\bm{g}_{ij}^\top \bm{g}_{ij'})^2}},
\end{equation}
where the last equality is due to ${^i\bm{g}}_{ij}^\top {^i\bm{g}}_{ij'}=(Q_i^\top \bm{g}_{ij})^\top Q_i^\top \bm{g}_{ij'}=\bm{g}_{ij}^\top I_3\bm{g}_{ij'} =\bm{g}_{ij}^\top \bm{g}_{ij'}$. 

\begin{figure}[htb!] 
    \centering
    \includegraphics[scale=1]{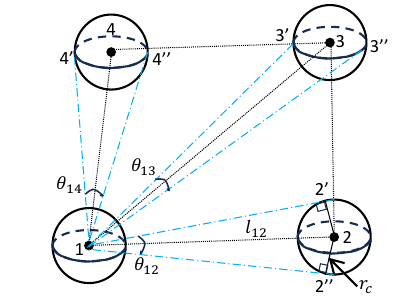}
    \caption{Elevation angles in 3D.}
    \label{f2}
\end{figure}

For the 3D case, each agent is equipped with a ball of radius $r_c$. The elevation angle $\theta_{ij} \in(0,\frac{\pi}{3})$ from agent $i$ to agent $j$ is defined as 
\begin{equation}\label{e8}
\theta_{ij} =\angle j'ij''=\arccos(\bm{g}_{ij'}^\top \bm{g}_{ij''})=2\angle j'ij= 2\arcsin(r_c/l_{ij}),
\end{equation}
where $j'$ and $j''$ are two distinct points on the surface of agent $j$'s ball such that $j,\ j',\ j'',\ i$ are coplanar and $\bm{g}_{ij'}^\top \bm{g}_{jj'}=0$, $\bm{g}_{ij''}^\top \bm{g}_{jj''}=0$ as shown in Figure \ref{f2}. Note that $\bm{g}_{ij'}^\top \bm{g}_{jj'}=0$ and $\bm{g}_{ij''}^\top \bm{g}_{jj''}=0$ as they are perpendicular vectors. The range of $\theta_{ij}$ is between 0 and $\frac{\pi}{3}$ because when the agent $i$' s ball touches the agent $j$' s ball, then $l_{ij} = 2r_c$, and the elevation angle becomes $ \theta_{ij} = 2\arcsin(\frac{r_c}{2r_c}) = \frac{\pi}{3}$. Figure \ref{f2} depicts three elevation angles $\theta_{12},\ \theta_{13},\ \theta_{14}$ measured from agent 1 in 3D. We define the elevation angle function $\bm{f_E}(\bm{p}):\mathbb{R}^{3n}\rightarrow \mathbb{R}^m$ in 3D as $\bm{f_E}(\bm{p}):=[f_1,...,f_k,...,f_m]^\top$, where 
\begin{align} \label{e9}
\text{\small $ f_k=f_{ij}=\text{cosec}\left(\frac{\theta_{ij}}{2}\right)
    =\frac{1}{\sin\left(\theta_{ij}/2\right)}= \frac{l_{ij}}{r_c}, \quad\ k = 1,...,m.$}
\end{align}
We can rewrite \eqref{e9} using local bearing vectors as
\begin{equation} \label{e10}
    f_{ij} {=}\frac{1}{\sqrt{(1-{^i\bm{g}}_{ij'}^\top {^i\bm{g}}_{ij''})/2}}
    = \frac{1}{\sqrt{(1-\bm{g}_{ij'}^\top \bm{g}_{ij''})/2}}.
\end{equation}

The desired formation is represented by $\bm{f_E}^*=[f_1^*,...,f_k^*,...,f_m^*]^\top$, indicating $f_{ij}^*=\frac{l_{ij}^*}{\rho}$, for all $(i,j)\in\mathcal{E}$. Here, $l_{ij}^*$ denotes the desired distance, and $\rho:=h_c$ for 2D and $\rho=r_c$ for 3D. The time derivative of the elevation angle function is 
\begin{equation} \label{e11}
    \frac{d \bm{f_E}}{dt}=\frac{\partial \bm{f_E}}{\partial \bm{p}}\dot{\bm{p}}=R_E(\bm{p})\dot{\bm{p}},
\end{equation}
where $R_E(\bm{p})\in \mathbb{R}^{m\times 3n}$ is the elevation angle rigidity matrix \cite{chen_2022_gradientbased}. The rigidity matrix $R_E$ can be expressed as 
\begin{equation} \label{e12}
    R_E =\frac{\partial \bm{f_E}}{\partial \bm{p}} = \text{diag}(\rho^{-1} \bm{g}_1^\top,..., \rho^{-1} \bm{g}_m^\top)(\bar{H}). 
\end{equation}
where $\bar{H} = H \otimes I_d$. Note that the rigidity matrix $R_E$ can be used in both 2D and 3D scenarios. We refer readers to \cite{chen_2022_gradientbased} for further details about $R_E$. \vspace{0.2cm}\\
\textbf{Lemma 1 \cite{chen_2022_gradientbased}.}\textit{ A framework $(\mathcal{G},\bm{p})$ is infinitesimally elevation angle rigid if and only if the desired formation satisfies} $\text{rank}(R_E)=dn-d(d+1)/2$. 

\subsection{Finite-time Input-to-State Stability}
The benchmark work in \cite{hong_2010_finite} presents a FTISS analysis for autonomous systems. Consider the following system\\
\begin{equation}\label{e13}
    \dot{\bm{z}}=f(\bm{z},\bm{v}),\quad f(\bm{0},\bm{0})=\bm{0},
\end{equation}
where $\bm{z}(t)\in\mathbb{R}^p$ is the state and $\bm{v}(t)\in\mathbb{R}^q$ is the input. A function $\varphi:\mathbb{R}^+\rightarrow\mathbb{R}^+$  is called a $K$-function if it is continuous, strictly increasing, and satisfies $\varphi(0)=0$. 
If $\varphi(s)$ is a $K$-function and $\varphi(s) \rightarrow \infty$ as $s \rightarrow \infty$, then it is a $K_\infty$-function. Moreover, a function $\vartheta:\mathbb{R}^+\times \mathbb{R}^+\rightarrow \mathbb{R}^+$ is called a generalized $KL$-function ($GKL$-function) if the mapping $s\mapsto\vartheta(s,0)$ is a $K$-function, and for each fixed $s\geq0$ the mapping $t\mapsto\vartheta(s,t)$ is continuous and decreases to zero as $t\rightarrow T(s)$ for $T(s)\in[0,\infty)$. The definition of FTISS is provided below.\vspace{0.2cm}\\
\textbf{Definition 1 \cite{hong_2010_finite}.} \textit{System \eqref{e13} is FTISS if there exists some neighborhoods $U\in\mathbb{R}^p$ of zero and $U_v\in\mathbb{R}^q$ of zero such that, for all $\bm{z}_0=\bm{z}(0)\in U$ and input $\bm{v}\in U_v$ that is measurable and bounded, each solution $Z(\bm{z}_0,\bm{v},t)$ is defined for $t\geq0$ and satisfies}
\begin{equation} \label{e14}
    \|Z(\bm{z}_0,\bm{v},t)\|\leq \vartheta(\|\bm{z}_0\|,t)+ \varphi(\|\bm{v}\|_\infty),
\end{equation}
\textit{where $\varphi$ is a $K$-function and $\vartheta$ is a $GKL$-function with $\vartheta(s,t)=0$ when $t\geq T(s)$ with $T(s)$ continuous with respect to $s$ and $T(0)=0$.\vspace{0.2cm}}

Suppose that there exist $K_\infty$-functions $\phi_1$ and $\phi_2$ such that
\begin{equation} \label{e15}
    \phi_1(\bm{z})\leq V(\bm{z})\leq \phi_2(\bm{z}),\quad \forall \bm{z}\in\mathbb{R}^d.
\end{equation}
Furthermore, there exists $K$-functions $\phi_3$ and $\phi_4$ such that for any solution $\bm{z}(t)$ and input $\bm{v}(t)$ of system \eqref{e13}, it holds 
\begin{equation}\label{e16}
    \|\bm{z}(t)\|\geq \phi_4(\|\bm{v}(t)\|) \Rightarrow \dot{V}(\bm{z}(t))|_{\eqref{e13}}\leq -\phi_3(\|\bm{z}\|),
\end{equation}
for $t\geq 0$. Then, the following result is obtained.\vspace{0.2cm}\\
\textbf{Lemma 2 \cite{hong_2010_finite}.} \textit{A continuous function $V(\bm{z})$ is called a FTISS-Lyapunov function for system \eqref{e13} if conditions \eqref{e15} and \eqref{e16} are satisfied, and $\phi_3(\|\bm{z}\|)\sim V(\bm{z})^\alpha$ for $\alpha\in(0,1)$. 
The system \eqref{e13} is FTISS with $\bm{v}$ as input if it has a FTISS-Lyapunov function.}\vspace{0.2cm}

Note that FTISS implies input-to-state stability. Moreover, FTISS implies finite-time stability when $\bm{v}=\bm{0}_q$.

\section{FTISS Formation Control with Unknown Constant Disturbance Rejection}
In this section, we first state the problem and then propose a FTISS bearing-only formation control law to address it.
\subsection{Problem Formulation}
Let $\mathcal{V}_l$ and $\mathcal{V}_f$ denote the sets of leader and follower vertices, respectively. Without loss of generality, assume that the multi-agent system has at least two leaders, with the first $n_l$ vertices in $\mathcal{V}$ being leaders and the remaining $n_f = n - n_l$ vertices are followers, such that $\mathcal{V}_l = \{1, \ldots, n_l\}$ and $\mathcal{V}_f = \{n_l + 1, \ldots, n\}$. We assume that the leaders are stationary, and this assumption will be relaxed in the next section. The dynamics of the agents is modeled as
\begin{equation} \label{e17} 
\begin{array}{ll} 
    ^i\dot{\bm{p}}_i = \bm{0}_3, & i\in\mathcal{V}_l,\\
    ^i\dot{\bm{p}}_i = {^i\bm{u}_i}+{^i\bm{\omega}_i},& i\in\mathcal{V}_f,  
\end{array}
\end{equation}
where $^i\bm{u}_i\in\mathbb{R}^3$ and $^i\bm{\omega}_i\in\mathbb{R}^3$ are the control input and the unknown constant disturbance in follower $i$'s local coordinate frame. The disturbance measured in the global coordinate frame is given by $\bm{\omega}_i=Q_i {^i\bm{\omega}}_i,\ i\in\mathcal{V}_f$. We assume that the following assumptions hold:\vspace{0.2cm}\\
\textit{ Assumption 1: The sensing topology among agents is expressed by an undirected graph $\mathcal{G(V,E)}$. Each follower $i\in \mathcal{V}_f$ can measure the local bearing vectors $^i\bm{g}_{ij}$ to its neighboring agents.}\vspace{0.2cm}\\
\textit{Assumption 2: The target formation is infinitesimally elevation angle rigid.}\vspace{0.2cm}\\
\textit{Assumption 3: Collision avoidance of agents is not considered, and the initial positions of agents are selected to ensure that no collision will occur.}\vspace{0.2cm}\\
We consider the formation control with constant disturbance problem as follows.\vspace{0.2cm}\\
\textbf{Problem 1.} \textit{Consider the leader-follower dynamics with constant disturbance modeled by \eqref{e17} and suppose Assumptions 1-3 hold. Design the control law $ ^i \bm{u}_i$ such that $ \bm{f_E} -\bm{f_E}^*\rightarrow \bm{0}_m$ as $t \rightarrow \infty$. Moreover, it should be guaranteed that the system under the proposed control law is FTISS.
}

\subsection{FTISS Bearing-only Formation Control Law}
For each follower $i$, we propose the following control law that rejects constant disturbance
\begin{equation} \label{e18}
\begin{split}
    ^i\bm{u}_i &= k_p\text{sig}\Bigl(\sum_{j\in\mathcal{N}_i} {^i\bm{g}_{ij}}(f_{ij}-f_{ij}^*)\Bigl)^\alpha -{^i\hat{\bm{\omega}}_i},\\
    ^i\dot{\hat{\bm{\omega}}}_i &= -k_e \sum_{j\in\mathcal{N}_i} {^i\bm{g}_{ij}} (f_{ij}-f_{ij}^*),\\
\end{split}    
\end{equation}
where $^i\hat{\bm{\omega}}_i\in\mathbb{R}^3$ is the estimate of the constant disturbance, $\alpha\in(0,1)$, and $k_p,\ k_e>0$ are positive control gains. We set the initial condition of $^i\hat{\bm{\omega}}_i$ to $\bm{0}_3$. Obviously, the control input \eqref{e18} can be obtained using only local bearing measurement, as explained in \eqref{e7} for 2D and \eqref{e10} for 3D. 
The follower dynamics in \eqref{e17} steered by the control law \eqref{e18} can be expressed in the global coordinate frame as
\begin{equation} \label{e19}
\begin{split}
    \dot{\bm{p}}_i &= k_pQ_i\text{sig}\Bigl(\sum_{j\in\mathcal{N}_i} Q_i^\top\bm{g}_{ij}(f_{ij}-f_{ij}^*)\Bigl)^\alpha -{\hat{\bm{\omega}}_i}+\bm{\omega}_i,\\
    \dot{\hat{\bm{\omega}}}_i &= -k_e \sum_{j\in\mathcal{N}_i} {\bm{g}_{ij}} (f_{ij}-f_{ij}^*),\quad i\in\mathcal{V}_f,
\end{split}    
\end{equation}
where $\hat{\bm{\omega}}_i=Q_i {^i\hat{\bm{\omega}}_i}$. The compact form of the follower dynamics \eqref{e19} and the leader dynamics can be written as
\begin{equation}\label{e20}
\begin{split}
    \dot{\bm{p}} &\stackrel{\eqref{e12}}{=}  -k_p Q\bar{M}\text{sig}\Bigl(\rho\bar{M}Q^\top R_E^\top (\bm{f_{E}}-\bm{f_{E}}^*)\Bigl)^\alpha -{\bar{M}\hat{\bm{\omega}}} +\bar{M}\bm{\omega}, \\
    \dot{\hat{\bm{\omega}}} &\stackrel{\eqref{e12}}{=} k_e\rho\bar{M} R_E^\top (\bm{f_{E}}-\bm{f_{E}}^*),
\end{split}
\end{equation}
where $\hat{\bm{\omega}}=[0,...,0,\hat{\bm{\omega}}_{n_l+1}^\top,...,\hat{\bm{\omega}}_n^\top]^\top\in\mathbb{R}^{3n}$, $\bm{\omega} = [0,...,0,\bm{\omega}_{n_l+1}^\top,...,\bm{\omega}_n^\top]^\top\in\mathbb{R}^{3n}$, $Q=\text{diag}(I_{3n_l},Q_{n_l+1},...,Q_n)$, $\bar{M} = M\otimes I_d\in \mathbb{R}^{3n\times 3n}$ with $M=\begin{bmatrix}0_{n_l\times n_l} & 0\\0 & I_{n_f} \end{bmatrix}$.

\subsection{Stability Analysis}
We define the formation error as $\bm{z}_e:=\bm{f_E}-\bm{f_E}^*$. Then, the error dynamics is given as 
\begin{align*}
    \dot{\bm{z}}_e&\stackrel{\eqref{e11}}{=}R_E\dot{\bm{p}}\\
    &\stackrel{\eqref{e20}}{=}  -k_p R_EQ\bar{M}\text{sig}\Bigl(\rho\bar{M}Q^\top R_E^\top \bm{z}_e\Bigl)^\alpha -R_E\bar{M}(\hat{\bm{\omega}} -\bm{\omega}). \numberthis\label{e21}
\end{align*} 

We first show the asymptotic stability of the system in Theorem 1 and then show the FTISS of the system in Theorem 2. \vspace{0.2cm}\\
\textbf{Theorem 1.} \textit{Consider the system \eqref{e17} driven by the control law \eqref{e18}. Under Assumptions 1-3, follower agents will achieve the desired formation shape, that is $\bm{z}_e\rightarrow \bm{0}_m$, as $t\rightarrow \infty$.}\\
\textit{Proof.} Let $\tilde{\bm{\omega}}:=\hat{\bm{\omega}}-\bm{\omega}$ and consider the following Lyapunov function
\begin{equation} \label{e22}
    V(\bm{z}_e,\tilde{\bm{\omega}}) = V_1 + \frac{1}{2k_e} \|\tilde{\bm{\omega}}\|^2,
\end{equation}
where $V_1(\bm{z}_e)= \frac{\rho}{2}\|\bm{z}_e\|^2$. Taking the time derivative of \eqref{e22}, we have
\begin{align*}
    \dot{V} &=\rho\bm{z}_e^\top \dot{\bm{z}}_e + \frac{1}{k_e}\tilde{\bm{\omega}}^\top \dot{\hat{\bm{\omega}}}\\
    &\stackrel{\eqref{e20},\eqref{e21}}{=} - k_p\rho \bm{z}_e^\top R_EQ\bar{M}\text{sig}\Bigl(\rho\bar{M}Q^\top R_E^\top \bm{z}_e\Bigl)^\alpha \\
    &\qquad - \rho\bm{z}_e^\top R_E  \bar{M}\tilde{\bm{\omega}}
    +\rho\tilde{\bm{\omega}}^\top\bar{M} R_E^\top \bm{z}_e \\
    &\stackrel{\eqref{signorm}}{=}- k_p\rho^{1+\alpha} \left|\bar{M}Q^\top R_E^\top \bm{z}_e\right|^{1+\alpha}\\
    &\stackrel{\eqref{e2}}{\leq} -k_p\rho^{1+\alpha} \left\|\bar{M}Q^\top R_E^\top \bm{z}_e\right\|^{1+\alpha}. \numberthis \label{e23}
\end{align*}
Since $\bar{M}=\bar{M}^\top=\bar{M}^2$, $\bar{M}Q^\top=Q^\top\bar{M}$ and $\|Q^\top \bm{x}\|=\sqrt{\bm{x}^\top Q Q^\top \bm{x}}=\|\bm{x}\|$, for $\bm{x}\in\mathbb{R}^{3n}$, \eqref{e23} can be further derived as
\begin{align*}
    \dot{V} &\leq -k_p\rho^{1+\alpha} \left\|\bar{M} R_E^\top \bm{z}_e\right\|^{1+\alpha}\\
    &\leq -k_p\rho^{1+\alpha}\Big(\lambda_{min}^+(R_E \bar{M}R_E^\top)\Big)^\frac{1+\alpha}{2} \|\bm{z}_e\|^{1+\alpha}, \numberthis \label{e24}
\end{align*}
where $\lambda_{min}^+(R_E \bar{M}R_E^\top)$ is the smallest positive eigenvalue of $R_E \bar{M}R_E^\top$. The last inequality is derived using Lemma 3 (provided in the Appendix). Hence, $\dot{V}$ is negative semidefinite. Using LaSalle's invariance principle, every solution of the system starting from a compact and positive invariant set containing $\bm{z}_e=\bm{0}_m$ converges to the largest invariant set where $\dot{V}=0$. From \eqref{e24}, we obtain $\bm{z}_e\rightarrow\bm{0}_m$ as $t\rightarrow \infty$. $\hfill\blacksquare$ \vspace{0.2cm} 

Theorem 1 shows that $\bm{z}_e=\bm{0}_m$ is asymptotically stable. Now, we will show that the system \eqref{e21} is FTISS.\vspace{0.2cm}\\
\textbf{Theorem 2.} \textit{Considering the same assumptions as in Theorem 1, the error dynamics \eqref{e21} is FTISS.}\\
\textit{Proof.} The time derivative of the scalar function $V_1$ is 
\begin{align*}
    \dot{V}_1 &\stackrel{\eqref{e21}}{=} - k_p\rho \bm{z}_e^\top R_EQ\bar{M}\text{sig}\Bigl(\rho\bar{M}Q^\top R_E^\top \bm{z}_e\Bigl)^\alpha \\
    &\qquad - \rho\bm{z}_e^\top R_E  \bar{M}\tilde{\bm{\omega}}\\
    &\stackrel{\eqref{e24}}{\leq} -k_p\rho^{1+\alpha}\Big(\lambda_{min}^+(R_E \bar{M}R_E^\top)\Big)^\frac{1+\alpha}{2} \|\bm{z}_e\|^{1+\alpha} \\
    &\qquad+ \|\bm{z}_e\| \|\rho R_E\bar{M}\|\|\tilde{\bm{\omega}}\|, \numberthis \label{e25}
\end{align*}
where the last inequality is obtained using Cauchy-Schwarz inequality. We can further derive \eqref{e25} as
\begin{align*}
    \dot{V}_1 &\stackrel{\eqref{e12}}{\leq}  -k_p\rho^{1+\alpha}\Big(\lambda_{min}^+(R_E \bar{M}R_E^\top)\Big)^\frac{1+\alpha}{2} \|\bm{z}_e\|^{1+\alpha} \\
    &\qquad +\|\bm{z}_e\| \|\text{diag}( \bm{g}_1^\top,..., \bm{g}_m^\top)\|\|\bar{H}\bar{M}\|\|\tilde{\bm{\omega}}\|\\
    &= -k_p\rho^{1+\alpha}\Big(\lambda_{min}^+(R_E \bar{M}R_E^\top)\Big)^\frac{1+\alpha}{2} \|\bm{z}_e\|^{1+\alpha}\\ &\qquad+\|\bm{z}_e\| \|\bar{H}\bar{M}\|\|\tilde{\bm{\omega}}\|, \numberthis\label{e26}
\end{align*}
where the last equality is due to $\|\text{diag}( \bm{g}_1^\top,..., \bm{g}_m^\top)\|=1$. For 
\begin{equation} \label{e27}
    \|\tilde{\bm{\omega}}\|\leq \frac{k_p\rho^{1+\alpha} \Big(\lambda_{min}^+(R_E \bar{M}R_E^\top)\Big)^\frac{1+\alpha}{2}}{2\|\bar{H}\bar{M}\|}\|\bm{z}_e\|^{\alpha},
\end{equation}
we obtain
\begin{align*}
    \dot{V}_1 &\leq -\frac{1}{2} k_p\rho^{1+\alpha}\Big(\lambda_{min}^+(R_E \bar{M}R_E^\top)\Big)^\frac{1+\alpha}{2} \|\bm{z}_e\|^{1+\alpha}\\
    &=- 2^{\frac{\alpha-1}{2}}k_p\rho^\frac{1+\alpha}{2}\Big(\lambda_{min}^+(R_E \bar{M}R_E^\top)\Big)^\frac{1+\alpha}{2} V_1^\frac{1+\alpha}{2}. \numberthis\label{e28}
\end{align*}
Since $V_1$ is radially unbounded, condition \eqref{e15} is satisfied. Note that $\frac{1+\alpha}{2}\in(0,1)$. By combining \eqref{e27} and \eqref{e28}, we conclude that $V_1$ serves as a FTISS Lyapunov function. Consequently, the error dynamics \eqref{e21} is FTISS according to Lemma 2. $\hfill \blacksquare$

By combining the results of Theorem 1 and Theorem 2, we conclude that the formation errors will converge to the neighborhood of the equilibrium in finite time and subsequently converge to the equilibrium. 

\section{FTISS Formation Tracking Control with Unknown Constant Moving Leaders}
In this section, we extend the findings of the previous section to account for leaders moving at a common constant velocity.
\subsection{Problem Formulation}
Assuming the leaders are moving at a common constant velocity, the leader-follower dynamics with moving leaders can be expressed as
\begin{equation}\label{e29}
\begin{array}{ll} 
    ^i\dot{\bm{p}}_i = Q_i^\top \bm{v}^*,& i \in\mathcal{V}_l,\\
    ^i\dot{\bm{p}}_i = {^i\bm{u}_i}+{^i\bm{\omega}_i},& i \in\mathcal{V}_f,
\end{array}  
\end{equation}
where $\bm{v}^*\in \mathbb{R}^3$ is the velocity of the leaders. For the leaders dynamics in global coordinate frame, we have $\dot{\bm{p}}_i = \bm{v}^*, i \in\mathcal{V}_l$. Equation \eqref{e29} in the global coordinate frame can be written in the following compact form
\begin{equation} \label{e30}
    \dot{\bm{p}} = \begin{bmatrix} \bm{0}_{3n_l} \\\dot{\bm{p}}_f-\bm{1}_{n_f}\otimes \bm{v}^* \end{bmatrix}+\bm{1}_n\otimes \bm{v}^* ,
\end{equation}
where $\dot{\bm{p}}_f$ denotes the overall dynamics of followers. Let $\tilde{\bm{p}}:=\bm{p}-\bm{p}^*$, where $\dot{\bm{p}}^*=\bm{1}_n\otimes \bm{v}^*$. The time derivative of $\tilde{\bm{p}}$ is
\begin{align*}
    \dot{\tilde{\bm{p}}} &= \begin{bmatrix} \bm{0}_{dn_l} \\\dot{\bm{p}}_f-\bm{1}_{n_f}\otimes \bm{v}^* \end{bmatrix}\\
    &\stackrel{\eqref{e20}}{=}  -k_p Q\bar{M}\text{sig}\Bigl(\rho\bar{M}Q^\top R_E^\top (\bm{f_{E}}-\bm{f_{E}}^*)\Bigl)^\alpha -{\bar{M}\hat{\bm{\omega}}}\\ 
    &\qquad +\bar{M}\bm{\omega}-\bar{M}(\bm{1}_n\otimes \bm{v}^*).\numberthis \label{e31}
\end{align*}
Compared to the previous problem, we now have an additional constant term, $-\bar{M}(\bm{1}_n\otimes \bm{v}^*)$. Under Assumptions 1-3, the following problem is introduced.\vspace{0.2cm}\\
\textbf{Problem 2.} \textit{Given that the vector of exogenous disturbances and the common velocity of the leaders are unknown, design a control law for the followers such that $\bm{z}_e \rightarrow \bm{0}_m$ as $t \rightarrow \infty$. Furthermore, ensure that $\bm{z}_e$ converges to the neighborhood of $\bm{z}_e=\bm{0}_m$ in finite time.}

\subsection{Stability Analysis}
The following theorem is presented to solve Problem 2.\vspace{0.2cm}\\
\textbf{Theorem 3.} \textit{Consider the dynamics \eqref{e29} and the control law \eqref{e18}. The elevation angle errors $\bm{z}_e\rightarrow \bm{0}_m$ as $t\rightarrow \infty$ under Assumptions 1-3. Moreover, the system under the proposed control law is FTISS.}\\
\textit{Proof.} The term $\bar{M}(\bm{\omega}-\bm{1}_n\otimes \bm{v}^*)$ in \eqref{e29} can be treated as a total constant disturbance by redefining $\tilde{\bm{\omega}}:=\hat{\bm{\omega}}-\bm{\omega}+\bm{1}_n\otimes \bm{v}^*$. Using the proof outlined in Theorem 1, we can demonstrate that $\bm{z}_e$ converges to zero as $t \rightarrow \infty$. Similarly, we can use Theorem 2 to show that $\bm{z}_e$ converges to the neighborhood of $\bm{z}_e = \bm{0}_m$ in finite time. Once the desired formation shape is achieved, the followers will track the common velocity of the leaders to maintain the formation.  $\hfill \blacksquare$

\section{Simulation and Experimental Results}
\begin{figure}[htb!]
    \centering
    \begin{subfigure}[htb!]{0.5\textwidth}
         \centering
         \includegraphics[scale=0.4]{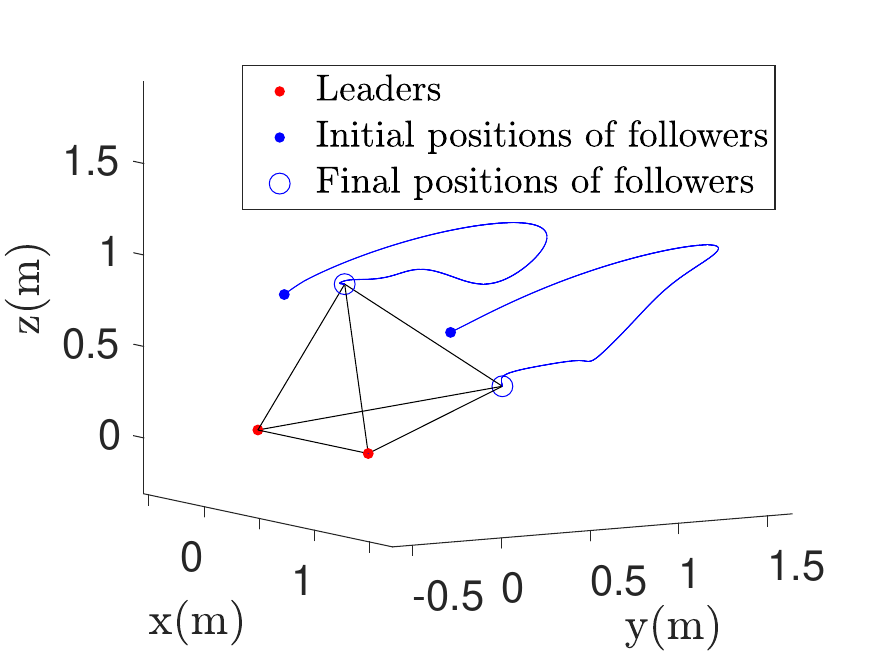}
         \caption{Agents trajectories.}
         \label{f3a}
     \end{subfigure}
     \begin{subfigure}[htb!]{0.5\textwidth}
         \centering
         \includegraphics[scale =0.4]{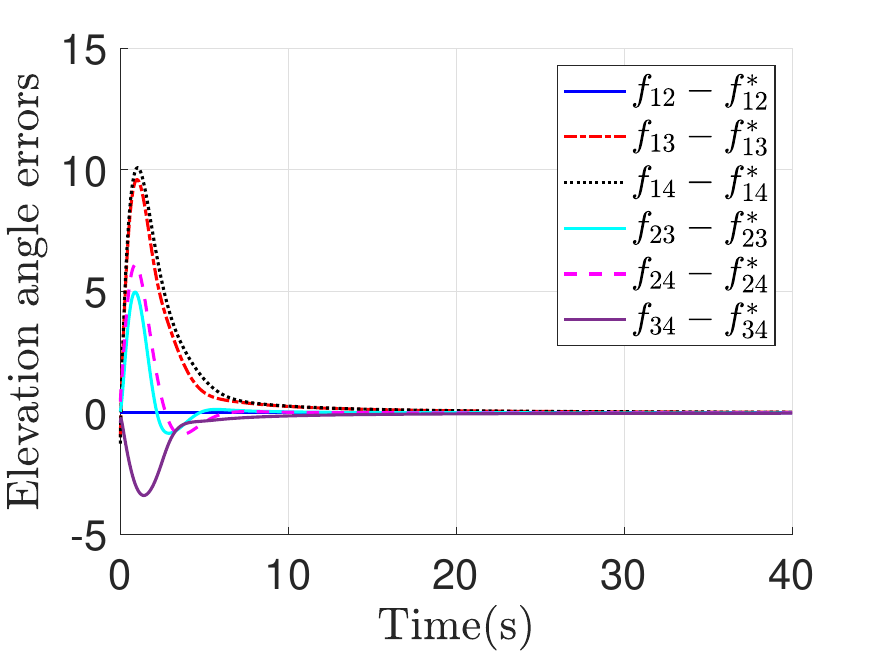}
         \caption{Elevation angle errors.}
         \label{f3b}
     \end{subfigure}
    \caption{FTISS bearing-only formation control with stationary leaders in 3D.}
    \label{f3}
\end{figure}

This section illustrates the effectiveness of the control law \eqref{e18}. 
Figure \ref{f3} shows simulation results for a multi-agent leader-follower system with two stationary leaders. The control parameters $k_p,\ k_e$ and $\alpha$ are set to $0.5,\ 0.1,$ and $0.5$, respectively. The initial positions of the leaders and followers are $\bm{p}_1(0) = [-0.5,0,0]^\top,\ \bm{p}_2(0) = [0.5,0,0]^\top,\ \bm{p}_3(0)=[-0.1,-0.1,0.8]^\top,$ and $ \bm{p}_4(0)=[-0.2,0.9,0.5]^\top$. The artificially generated disturbance experienced by each follower is set to $\bm{\omega}_i= [2,2,2]^\top$. Note that the value of this disturbance is unknown to agents. The desired formation is a tetrahedron. As shown in Figure \ref{f2}, the elevation angle errors converge to zero, indicating that the desired formation is achieved even in the presence of unknown constant disturbance. 

\begin{figure}[htb!]
    \centering
    \begin{subfigure}[htb!]{0.5\textwidth}
         \centering
         \includegraphics[scale=0.4]{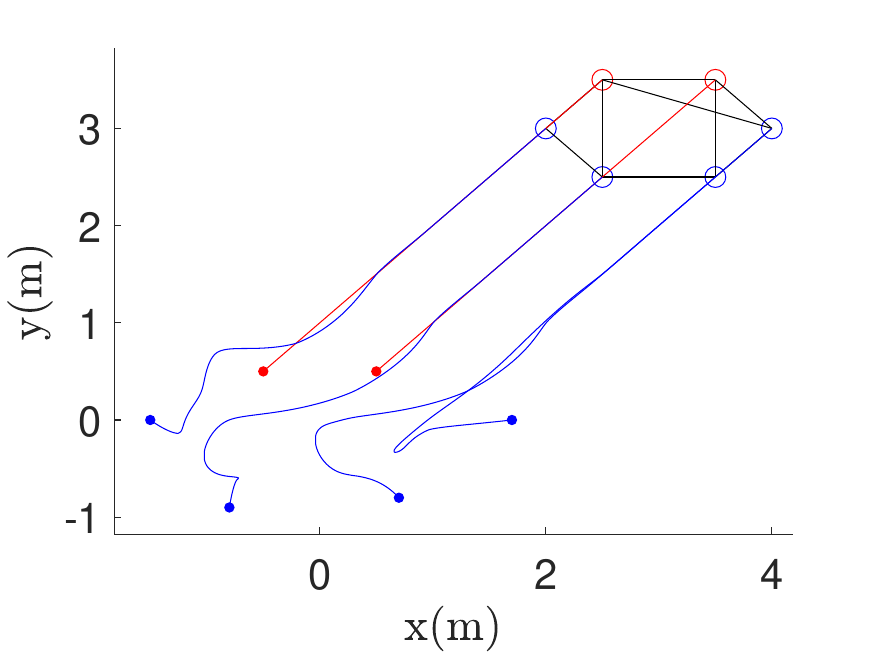}
         \caption{Agents trajectories. Red dots: initial positions of the leaders. Blue dots: initial positions of the followers. Red circles: positions of the leaders at $t=30s$. Blue circles: positions of the followers at $t=30s$.}
         \label{f4a}
     \end{subfigure}
     \begin{subfigure}[htb!]{0.5\textwidth}
         \centering
         \includegraphics[scale =0.4]{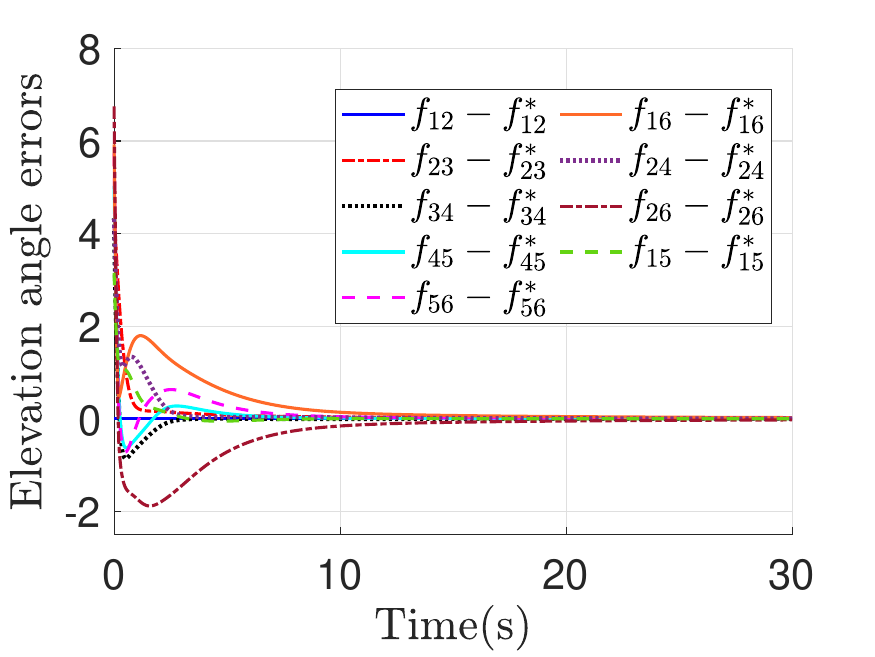}
         \caption{Elevation angle errors.}
         \label{f4b}
     \end{subfigure}
    \caption{FTISS bearing-only formation tracking control with moving leaders in 2D.}
    \label{f4}
\end{figure}

Figure \ref{f4} shows simulation results of the leader-follower dynamics with moving leaders \eqref{e29} steered by the same control law \eqref{e18} in 2D. The control parameters are set to $k_p = 1$, $k_e = 0.5$, and $\alpha = 0.5$. The initial positions of the leaders and followers are $\bm{p}_1(0) = [0.5,0.5,0]^\top,\ \bm{p}_2(0) = [-0.5,0.5,0]^\top,\ \bm{p}_3(0)=[-1.5,0,0]^\top$,\ $\bm{p}_4(0)=[-0.8,-0.9,0]^\top,\ \bm{p}_5(0)=[0.7,-0.8,0]^\top,$ and $\bm{p}_6(0)=[1.7,0,0]^\top$. The exogenous disturbance for each follower is set to $\bm{\omega}_i=[-1,-1,0]^\top$. The desired formation is a hexagon and the desired velocity of the leaders is set to $\bm{v}^* = [0.1,0.1,0]^\top$. As observed in Figure \ref{f4}, followers achieve the desired formation and track the moving leaders.

\begin{figure}[htb!]
    \centering
    \includegraphics[scale=0.4]{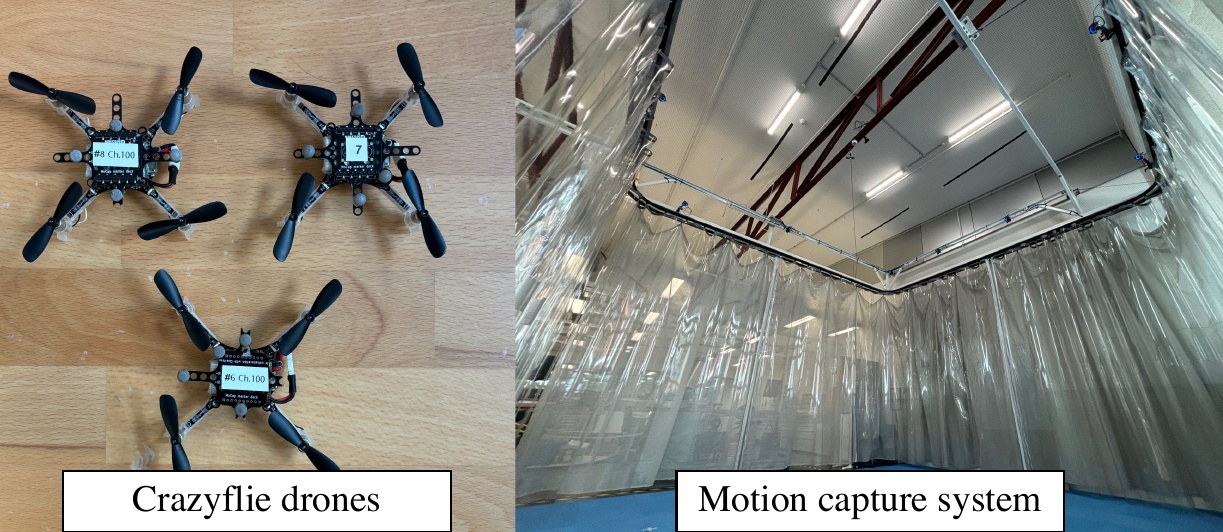}
    \caption{Experiment setup.}
    \label{f5}
\end{figure}

\begin{figure}[htb!]
    \centering
    \begin{subfigure}[htb!]{0.5\textwidth}
         \centering
         \includegraphics[scale=0.4]{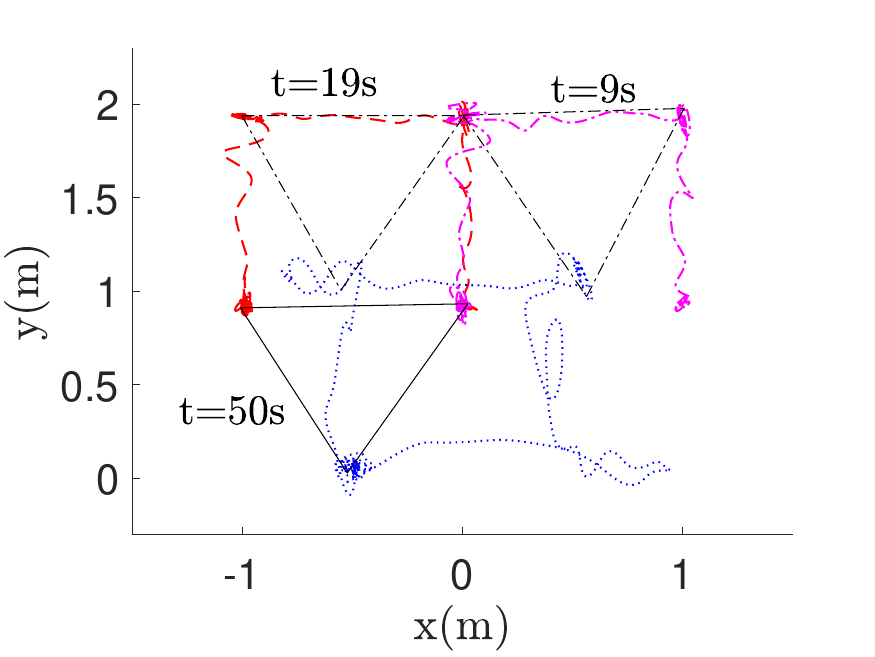}
         \caption{Agents trajectories. Magenta 	dash-dotted line: trajectory of the first leader. Red dashed line: trajectory of the second leader. Blue dotted line: trajectory of the follower.}
         \label{f6a}
     \end{subfigure}
     \begin{subfigure}[htb!]{0.5\textwidth}
         \centering
         \includegraphics[scale =0.4]{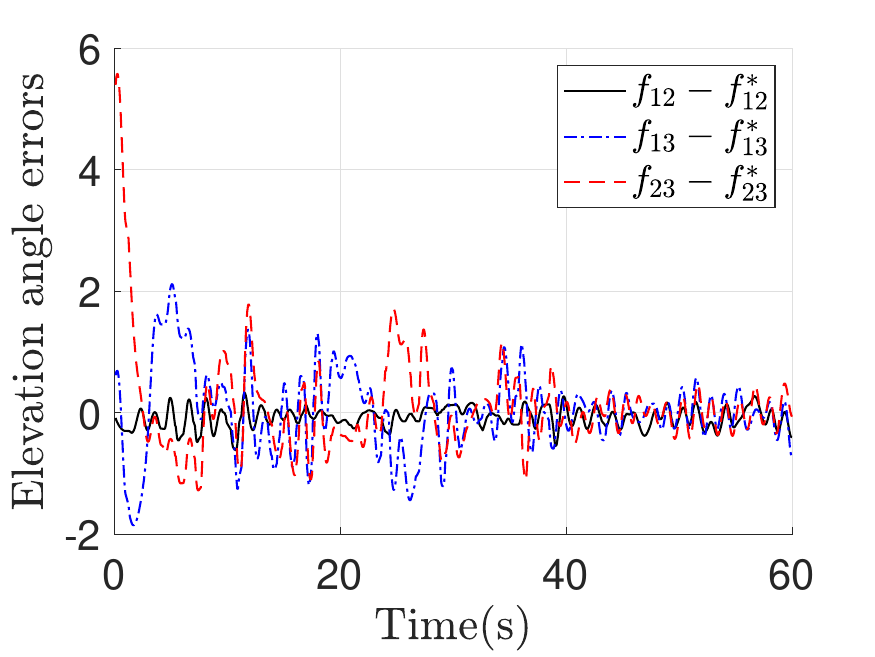}
         \caption{Elevation angle errors.}
         \label{f6b}
     \end{subfigure}
    \caption{Experiment of FTISS bearing-only formation control with moving leaders.}
    \label{f6}
\end{figure}

In addition to the simulation results, we have also implemented the control law \eqref{e18} on real quadcopters. The experiments were carried out using an Optitrack motion capture system including 10 Prime$^\text{x}$ 13 cameras and Crazyflie 2.1 drones, as shown in Figure \ref{f5}. The positions of the drones were sent to a computer in real time, which then calculated the elevation angles and relayed this information to the drones. We assume that drones are attached to virtual rods with $h_c = 0.15m$. The control parameters are set to $k_p=0.2$, $k_e=0.1$ and $\alpha=0.5$. The desired formation is an equilateral triangle. During the experiment, the leaders are controlled to move at a constant, identical velocity, changing direction at $t = 9s$ and $t = 19s$. As shown in Figure \ref{f6}, the follower agent successfully tracks the velocity of the leaders after achieving the desired formation, even when the leaders change their moving direction. It should be noted that the steady-state errors in Figure \ref{f6}(b) are attributable to several practical problems. Firstly, despite the leaders' velocity being programmed to remain constant, Figure \ref{f6}(a) indicates that it is not constant, probably due to unavoidable and varying disturbances in the experimental environment. Moreover, time delays resulting from vision processing and data transmission also contributed to these errors. Nevertheless, the overall trend in Figure \ref{f6}(b) indicates a successful convergence of the formation. The experiment video is available at: \url{https://youtu.be/68p3OsggEOE}.

\section{Conclusion}
This paper proposes a FTISS bearing-only formation tracking control law, that effectively rejects unknown constant disturbances without the need for a global coordinate frame. The proposed control law ensures that the formation error converges to the neighborhood of zero in a finite time and subsequently to zero asymptotically. Furthermore, we demonstrate that the same control law can be applied to leaders moving at a constant velocity. Future work could explore formation control of more complex agent dynamics in the presence of disturbances.

\section*{Appendix}
\textbf{Lemma 3 \cite{cheah_2024_bearingbased}.} \textit{Consider a vector $\bm{x}= [0,...,0,x_k,...,x_m]^\top\in\mathbb{R}^m$, with the number of the first zero rows depending on the number of edges connecting the leaders. Under Assumption 2, $\bm{x}^\top R_E\bar{M}R_E^\top \bm{x} \geq \lambda_{min}^+ (R_E\bar{M}R_E^\top)\|\bm{x}\|^2$, where $\lambda_{min}^+ (R_E\bar{M}R_E^\top)$ is the smallest positive eigenvalue of $R_E\bar{M}R_E^\top$.}

\bibliographystyle{named.bst}
\bibliography{references.bib}

\begin{thebibliography}{}

\bibitem[\protect\citeauthoryear{Ahn}{2020}]{Ahn_2020_formation}
Hyo-Sung Ahn.
\newblock {\em Formation Control: Approaches for Distributed Agents}.
\newblock Springer, Cham, Germany, 2020.

\bibitem[\protect\citeauthoryear{Bae \bgroup \em et al.\egroup }{2022}]{bae_2022_distributed}
Yoo-Bin Bae, Seong-Ho Kwon, Young-Hun Lim, and Hyo-Sung Ahn.
\newblock Distributed bearing‐based formation control and network localization with exogenous disturbances.
\newblock {\em International Journal of Robust and Nonlinear Control}, 32, 05 2022.

\bibitem[\protect\citeauthoryear{Bhat and Bernstein}{2000}]{bhat_2000_finitetime}
Sanjay~P. Bhat and Dennis~S. Bernstein.
\newblock Finite-time stability of continuous autonomous systems.
\newblock {\em SIAM Journal on Control and Optimization}, 38:751--766, 01 2000.

\bibitem[\protect\citeauthoryear{Bishop \bgroup \em et al.\egroup }{2015}]{Bishop_2015_distributed}
Adrian~N. Bishop, Mohammad Deghat, Brian D.~O. Anderson, and Yiguang Hong.
\newblock Distributed formation control with relaxed motion requirements.
\newblock {\em International Journal of Robust and Nonlinear Control}, 25(17):3210--3230, 2015.

\bibitem[\protect\citeauthoryear{Cheah and Deghat}{2024}]{cheah_2024_bearingbased}
Hong~Liang Cheah and Mohammad Deghat.
\newblock Bearing-based leader-follower formation tracking control using elevation angle.
\newblock {\em IEEE Access}, 2024.

\bibitem[\protect\citeauthoryear{Chen and Cao}{2023}]{chen_2023_angle}
Liangming Chen and Ming Cao.
\newblock Angle rigidity for multiagent formations in 3-d.
\newblock {\em IEEE Transactions on Automatic Control}, 68(10):6130--6145, 2023.

\bibitem[\protect\citeauthoryear{Chen and Sun}{2022}]{chen_2022_gradientbased}
Liangming Chen and Zhiyong Sun.
\newblock Gradient-based bearing-only formation control: An elevation angle approach.
\newblock {\em Automatica}, 141:110310, 07 2022.

\bibitem[\protect\citeauthoryear{Chen \bgroup \em et al.\egroup }{2021}]{chen_2021_angle}
Liangming Chen, Ming Cao, and Chuanjiang Li.
\newblock Angle rigidity and its usage to stabilize multiagent formations in 2-d.
\newblock {\em IEEE Transactions on Automatic Control}, 66(8):3667--3681, 2021.

\bibitem[\protect\citeauthoryear{Chen \bgroup \em et al.\egroup }{2022}]{chen_2022_fuzzy}
Zitao Chen, Qin Wang, Enci Wang, and Mingsong Du.
\newblock Fuzzy adaptive formation control for a class of nonlinear systems with bearing-only measurements.
\newblock In {\em 41st Chinese Control Conference (CCC)}, page 4532–4537, 2022.

\bibitem[\protect\citeauthoryear{Chen \bgroup \em et al.\egroup }{2024}]{chen_2024_air}
Liangming Chen, Jiaping Xiao, Clarence Wei~Rui Teo, Jianqing Li, and Mir Feroskhan.
\newblock Air-ground collaborative control for angle-specified heterogeneous formations.
\newblock {\em IEEE Transactions on Intelligent Vehicles}, pages 1--13, 2024.

\bibitem[\protect\citeauthoryear{Cheng and Huang}{2024}]{CHENG_2024_adaptive}
Haoshu Cheng and Jie Huang.
\newblock Adaptive robust bearing-based formation control for multi-agent systems.
\newblock {\em Automatica}, 162:111509, 2024.

\bibitem[\protect\citeauthoryear{Ding \bgroup \em et al.\egroup }{2023}]{ding_2023_event}
Can Ding, Zhe Zhang, Jing Zhang, and Yingjie Zhang.
\newblock Event-triggered based bearing-only formation control for nonlinear multi-agent with unknown disturbance.
\newblock {\em Asian Journal of Control}, 2023.

\bibitem[\protect\citeauthoryear{Dung and Trinh}{2021}]{VanVu_2021_decentralized}
Van~Vu Dung and Minh~Hoang Trinh.
\newblock Decentralized sliding-mode control laws for the bearing-based formation tracking problem.
\newblock In {\em 2021 International Conference on Control, Automation and Information Sciences (ICCAIS)}, pages 67--72, 2021.

\bibitem[\protect\citeauthoryear{Garanayak and Mukherjee}{2023}]{Garanayak_2023_Bearingdisturbance}
Chinmay Garanayak and Dwaipayan Mukherjee.
\newblock Bearing-only formation control with bounded disturbances in agents’ local coordinate frames.
\newblock {\em IEEE Control Systems Letters}, 7:2940--2945, 2023.

\bibitem[\protect\citeauthoryear{Hong \bgroup \em et al.\egroup }{2010}]{hong_2010_finite}
Yiguang Hong, Zhong-Ping Jiang, and Gang Feng.
\newblock Finite-time input-to-state stability and applications to finite-time control design.
\newblock {\em SIAM Journal on Control and Optimization}, 48(7):4395--4418, 2010.

\bibitem[\protect\citeauthoryear{Jing \bgroup \em et al.\egroup }{2019}]{Jing_2019_angle}
Gangshan Jing, Guofeng Zhang, Heung Wing~Joseph Lee, and Long Wang.
\newblock Angle-based shape determination theory of planar graphs with application to formation stabilization.
\newblock {\em Automatica}, 105:117--129, 2019.

\bibitem[\protect\citeauthoryear{Mao \bgroup \em et al.\egroup }{2007}]{mao_2007_wireless}
Guoqiang Mao, Barış Fidan, and Brian~D.O. Anderson.
\newblock Wireless sensor network localization techniques.
\newblock {\em Computer Networks}, 51:2529--2553, 07 2007.

\bibitem[\protect\citeauthoryear{Mesbahi and Egerstedt}{2010}]{Mesbahi_2010_graph}
Mehran Mesbahi and Magnus Egerstedt.
\newblock {\em Graph Theoretic Methods in Multiagent Networks}.
\newblock Princeton Univ. Press, 2010.

\bibitem[\protect\citeauthoryear{Nguyen \bgroup \em et al.\egroup }{2024}]{nguyen_2023_bearingconstrained}
Thanh~Truong Nguyen, Dung~Van Vu, Tuynh~Van Pham, and Minh~Hoang Trinh.
\newblock Bearing-constrained leader-follower formation of single-integrators with disturbance rejection: Adaptive variable-structure approaches.
\newblock {\em IEEE Transactions on Cybernetics}, pages 1--14, 2024.

\bibitem[\protect\citeauthoryear{Song \bgroup \em et al.\egroup }{2024}]{song_2024_bearing}
Zilong Song, Miaomiao Xie, and Haocai Huang.
\newblock Bearing-only formation tracking control for multi-agent systems with time-varying velocity leaders.
\newblock {\em IEEE Control Systems Letters}, pages 1--1, 2024.

\bibitem[\protect\citeauthoryear{Su \bgroup \em et al.\egroup }{2024}]{Su_2024_bearing}
Haifan Su, Shanying Zhu, Cailian Chen, Ziwen Yang, and Xinping Guan.
\newblock Bearing-based robust formation tracking control of underactuated auvs with optimal parameter tuning.
\newblock {\em IEEE Transactions on Cybernetics}, pages 1--14, 2024.

\bibitem[\protect\citeauthoryear{Trinh \bgroup \em et al.\egroup }{2021}]{trinh_2021_robust}
Minh~Hoang Trinh, Quoc~Van Tran, Dung~Van Vu, Phuoc~Doan Nguyen, and Hyo-Sung Ahn.
\newblock Robust tracking control of bearing-constrained leader–follower formation.
\newblock {\em Automatica}, 131:109733, 09 2021.

\bibitem[\protect\citeauthoryear{Trinh \bgroup \em et al.\egroup }{2022}]{trinh_2022_adaptive}
Minh~Hoang Trinh, Thanh~Truong Nguyen, and Zhiyong Sun.
\newblock Adaptive distance-based formation control.
\newblock In {\em 2022 17th International Conference on Control, Automation, Robotics and Vision (ICARCV)}, pages 746--751, 2022.

\bibitem[\protect\citeauthoryear{Tron \bgroup \em et al.\egroup }{2016}]{tron_2016_a}
Roberto Tron, Justin Thomas, Giuseppe Loianno, Kostas Daniilidis, and Vijay Kumar.
\newblock A distributed optimization framework for localization and formation control: Applications to vision-based measurements.
\newblock {\em IEEE Control Systems Magazine}, 36:22--44, 08 2016.

\bibitem[\protect\citeauthoryear{Van~Tran \bgroup \em et al.\egroup }{2019}]{vantran_2019_finitetime}
Quoc Van~Tran, Minh~Hoang Trinh, Daniel Zelazo, Dwaipayan Mukherjee, and Hyo-Sung Ahn.
\newblock Finite-time bearing-only formation control via distributed global orientation estimation.
\newblock {\em IEEE Transactions on Control of Network Systems}, 6:702--712, 06 2019.

\bibitem[\protect\citeauthoryear{Van~Tran \bgroup \em et al.\egroup }{2023}]{vantran_2023_robust}
Quoc Van~Tran, Changyu Lee, Jinwhan Kim, and Hoang~Quang Nguyen.
\newblock Robust bearing-based formation tracking control of underactuated surface vessels: An output regulation approach.
\newblock {\em IEEE Transactions on Control of Network Systems}, 10(4):2048--2059, 2023.

\bibitem[\protect\citeauthoryear{Van~Vu \bgroup \em et al.\egroup }{2021}]{VanVu_2021_distance}
Dung Van~Vu, Minh~Hoang Trinh, Phuoc~Doan Nguyen, and Hyo-Sung Ahn.
\newblock Distance-based formation control with bounded disturbances.
\newblock {\em IEEE Control Systems Letters}, 5(2):451--456, 2021.

\bibitem[\protect\citeauthoryear{Wang and Liu}{2023}]{wang_2023_bearing}
Yujie Wang and Shuai Liu.
\newblock Bearing-based formation control simultaneously involving several heterogeneous multi-agent systems with nonlinear uncertainties.
\newblock In {\em 2023 62nd IEEE Conference on Decision and Control (CDC)}, pages 4429--4434, 2023.

\bibitem[\protect\citeauthoryear{Wang \bgroup \em et al.\egroup }{2022}]{wang_2022_adaptive}
Kun Wang, Tao Meng, Chengjin Yin, Heng Li, and Zhonghe Jin.
\newblock Adaptive bearing-based spacecraft formation control.
\newblock In {\em 2022 China Automation Congress (CAC)}, page 1850–1855, 2022.

\bibitem[\protect\citeauthoryear{Wu \bgroup \em et al.\egroup }{2023}]{wu_2023_finitetime}
Kefan Wu, Junyan Hu, Zhengtao Ding, and Farshad Arvin.
\newblock Finite-time fault-tolerant formation control for distributed multi-vehicle networks with bearing measurements.
\newblock {\em IEEE Transactions on Automation Science and Engineering}, page 1–12, 2023.

\bibitem[\protect\citeauthoryear{Wu \bgroup \em et al.\egroup }{2024}]{Wu_2024_distributed}
Kefan Wu, Junyan Hu, Zhenhong Li, Zhengtao Ding, and Farshad Arvin.
\newblock Distributed collision-free bearing coordination of multi-uav systems with actuator faults and time delays.
\newblock {\em IEEE Transactions on Intelligent Transportation Systems}, pages 1--14, 2024.

\bibitem[\protect\citeauthoryear{Xu \bgroup \em et al.\egroup }{2023}]{xu_2023_bearingbased}
Chuang Xu, Daniel Zelazo, and Baolin Wu.
\newblock Bearing-based formation control of second-order multiagent systems with bounded disturbances.
\newblock {\em International Journal of Robust and Nonlinear Control}, 08 2023.

\bibitem[\protect\citeauthoryear{Zhao and Zelazo}{2015}]{zhao_2015_maneuver}
Shiyu Zhao and Daniel Zelazo.
\newblock Bearing-based formation maneuvering.
\newblock In {\em IEEE International Symposium on Intelligent Control (ISIC)}, pages 658--663, 2015.

\bibitem[\protect\citeauthoryear{Zhao and Zelazo}{2016a}]{zhao_2016_bearing}
Shiyu Zhao and Daniel Zelazo.
\newblock Bearing rigidity and almost global bearing-only formation stabilization.
\newblock {\em IEEE Transactions on Automatic Control}, 61:1255--1268, 05 2016.

\bibitem[\protect\citeauthoryear{Zhao and Zelazo}{2016b}]{zhao_2016_localizability}
Shiyu Zhao and Daniel Zelazo.
\newblock Localizability and distributed protocols for bearing-based network localization in arbitrary dimensions.
\newblock {\em Automatica}, 69:334–341, 07 2016.

\bibitem[\protect\citeauthoryear{Zhao and Zelazo}{2017}]{zhao_2017_translational}
Shiyu Zhao and Daniel Zelazo.
\newblock Translational and scaling formation maneuver control via a bearing-based approach.
\newblock {\em IEEE Transactions on Control of Network Systems}, 4:429--438, 09 2017.

\bibitem[\protect\citeauthoryear{Zuo and Tie}{2014}]{zuo_2014_anew}
Zongyu Zuo and Lin Tie.
\newblock A new class of finite-time nonlinear consensus protocols for multi-agent systems.
\newblock {\em International Journal of Control}, 87(2):363--370, 2014.

\end{thebibliography}

\end{document}